\documentclass{article}
\usepackage{graphicx} 
\usepackage{titlesec}
\usepackage{float}
\usepackage{url}
\usepackage{hyperref}

\makeatletter

\renewenvironment{abstract}{%
  \if@twocolumn
    \section*{\abstractname}%
  \else
    \small
    \begin{center}%
      {\bfseries \vspace{-2em}}
    \end{center}%
    \quotation
  \fi}
  {\if@twocolumn\else\endquotation\fi}
\makeatother

\title{Psychology of Artificial Intelligence: Epistemological Markers of the Cognitive Analysis of Neural Networks}
\author{%
  \textbf{Michael Pichat} \\ 
  Cabinet Chrysippe R\&D (Neocognition), Paris University \\ 
  \href{mailto:michael.pichat@chrysippe.org}{michael.pichat@chrysippe.org} 
}

\begin{document}

\maketitle
\begin{abstract}
\noindent
    What is the "nature" of the cognitive processes and contents of an artificial neural network? In other words, how does an artificial intelligence fundamentally "think," and in what form does its knowledge reside? The psychology of artificial intelligence, as predicted by Asimov (1950), aims to study this AI probing and explainability-sensitive matter. This study requires a neuronal level of cognitive granularity, so as not to be limited solely to the secondary macro-cognitive results (such as cognitive and cultural biases) of synthetic neural cognition. A prerequisite for examining the latter is to clarify some epistemological milestones regarding the cognitive status we can attribute to its phenomenology.
\end{abstract}

\section*{What are the elementary cognitive building blocks of a neural network?}

{A layer of a neural network operates in a vector space whose dimensions can be associated with epistemological characteristics specific to that layer.

From a qualitative point of view, these dimensional characteristics can be linguistic (phonemic, phonetic, morphological, syntactic, lexical, semantic, pragmatic, etc.), visual (hue, saturation, luminosity, contrast, dimensional depth, shape, resolution, color depth, color spectrum, sharpness, noise, texture, contour, composition, scale, proportion, etc.), auditory (frequency, intensity, duration, timbre, pitch, melody, rhythm, harmony, noise, etc.), logical, contextual (relative position, etc.), qualitative, quantitative, etc. But these characteristics can also be of any other imaginable nature, whether "human-like" (i.e., characteristics referring to categories of thought and terms that humans possess) or "alien-like" (i.e., characteristics referring to categories of thought that humans do not currently possess) (Bills, 2023).

From a typological point of view, there can be no stable and exhaustive classification of these dimensional characteristics. They are a direct function of the nature of the data with which it is decided to feed a given neural network and the decreed modalities of its learning system, including the specific type of feedback that will be administered to it in the case of (totally or partially) supervised learning. Moreover, these characteristics are the immediate result of the specific architecture allocated to the neural network, its number of parameters and their nature, and the nature of its constitutive mathematical operators.

In the sense of formal logic, these dimensional characteristics can be associated with arguments and predicates (property, relation, transformation) or combinations thereof. These synthetic categories of thought are contingent cognitive constructions and not ontological elements. They pertain to an infinity of different possible cognitive modalities for segmenting the world.

We will return to these last two points after further clarifying some epistemological characteristics of synthetic neural cognition.

\section*{What is the "nature" of the cognitive activity operated by a neural layer?}

The input vector space of a neural layer is expressed in dimensional characteristics specific to that layer. The embedding of an input to this layer is thus formatted in this singular vector space; just as the weight matrix constitutive of this layer is calibrated in this same vector space.

Let us immediately avoid an elementary trap of cognitive anthropomorphism in the psychology of artificial intelligence by taking the case of a language model. A neural layer processing an incoming token does not reason about this token as such (like our human cognitive system, or at least our impression of it), does not "see" this token, but performs mathematical processing on the embedding in which this token is encoded. The epistemological analysis of this mathematical processing teaches us a series of lessons about the "nature" of the neural cognitive activity that is \textit{de facto} carried out by these mathematical operations.

In a given layer, the aggregation function of each formal neuron performs a well-determined mathematical processing on the embedding of an incoming token. For a given neuron \( i \), this processing is generated by the (horizontal) vector associated with it, which contains the weights \( W_{i,j} \) specific to it. Each of these weights \( W_{i,j} \) acts on one of the categorical dimensions \( j \) of the vector space in which the incoming tokens are expressed (see figure 1).

\begin{figure}[H]
    \centering
    \includegraphics[width=1\linewidth]{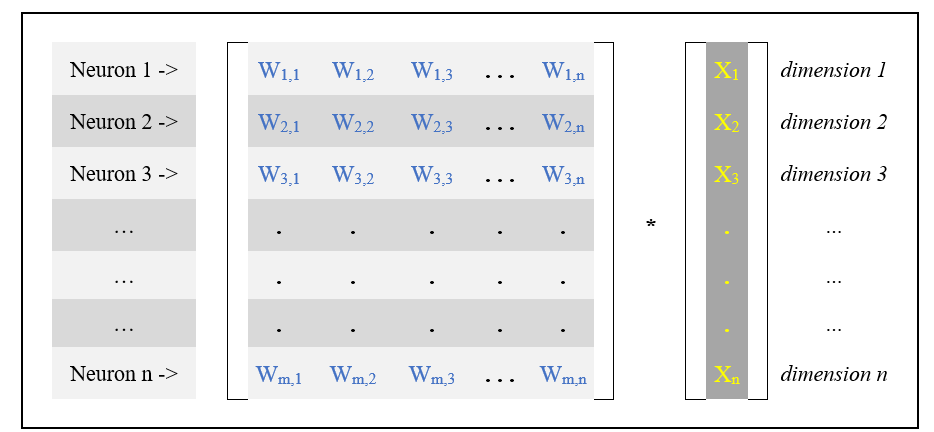}
    \caption{\textit{Matrix product of neural weights at the input of a layer.}}
    \label{fig:1}
\end{figure}

More precisely, each of these weights \( W_{i,j} \) will multiply a categorical dimension \( j \) of the incoming vector space. This multiplication of the weight \( W_{i,j} \) thus performs a cognitive activity of determining the intensity of attentional focus that neuron \( i \) will operate on this dimension \( j \). In other words, a weight is an epistemological selector that decides the level of importance (none, low, medium, high, total) to be given to a given dimensional characteristic. The resulting value of the performed multiplication indicates the (multiplicative) combination of two intensities: (i) the intensity of possession (\( X_j \)) by the incoming token of the categorical dimension \( j \), weighted by the attentional intensity \( W_{i,j} \) to be given to this categorical possession intensity. This value thus expresses the following information: what is the importance of the level of possession (by the token) of the categorical dimension? As such, we can cognitively qualify it as an epistemological residue expressing a weighted level of possession of an epistemological category of world segmentation, for a given step vector space (of progressive categorical segmentation) (see Figure 2).

\begin{figure}[H]
    \centering
    \includegraphics[width=1\linewidth]{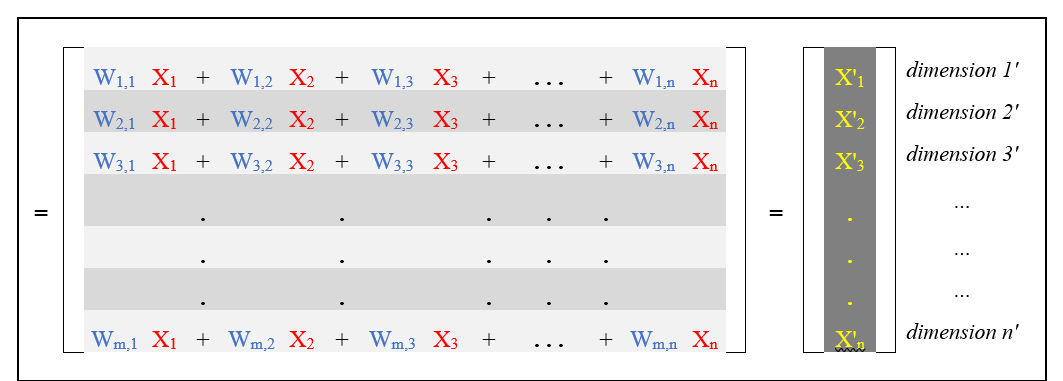}
    \caption{\textit{Result of the matrix product of neural weights at the output of a layer.}}
    \label{fig:2}
\end{figure}

Then, in the second step, all the products (attentional weight x categorical dimension) are added. This addition operationalizes \textit{de facto} a weighted epistemological fusion activity of the levels of possession, by the input token, of the categorical dimensions of the input vector space of the neural layer involved. This epistemological fusion thus constructs a new dimension \( j' \) of representation of the input token, a new categorical segmentation, and a new, more abstract categorical dimension. This cognitive fusion proceeds by a linear combination of all the epistemological residues of this initially formatted token in the dimensions specific to the input vector space \( j \): each new token expression abstraction is thus elaborated by selective concatenation of all the initial categorical dimensional segments.

Since the mentioned composition is additive, for a given neuron \( i \), the more intensely a token possesses the different starting dimensions \( j \) on which this neuron attentively focuses (i.e., for which the weights \( W_{i,j} \) of this neuron have high values), the more intensely the resulting new abstraction (the dimension \( j \)') will be possessed by the output embedding of this token. In other words, like the union operator in fuzzy logic, artificial neural abstraction proceeds by selective epistemological concatenation: a new output abstraction being the chosen union of certain characteristics distributed in the starting feature space (of characteristics), the "meaning" of a strong possession of this new abstraction is thus that of the simultaneous intense possession of all the initial superimposed dimensions of which it is the result of selective composition.

Thus, in part, the progressive abstraction cognitive process operated by a neural network works this way: at the end of each neural processing layer, a token is re-expressed in a new output vector space, each new dimension being the result of a "selective reduction" constitutive of this new dimension, of all the input dimensions. Each of these new, more abstract dimensions is characterized by its own mode of selectively combining (i.e., in a weighted manner) all the initial categorical dimensional segments of the starting vector space, that is, by giving more or less importance to each of these dimensional categories of world (of tokens) segmentation.

Such is the cognitive activity carried by the matrix products performed by the successive layers of a neural network: progressively projecting the embeddings of incoming tokens into increasingly abstract vector spaces, each new level of output abstraction achieved by weighted epistemological fusion of the input abstractions. From layer to layer, the embeddings of tokens are thus reformatted to express increasingly ethereal and fine-grained dimensions of abstraction (abstraction of abstraction, etc.).

\section*{What is the cognitive function of successive neural layers?}

The theory of conceptualization developed by Vergnaud (2016) offers us a valuable framework for thinking about the cognitive function of formal neurons in a neural network. According to the author, the primary form of knowledge is its operational one. The latter consists of knowledge-in-action, progressively fabricated through learning. This knowledge is constructed in the contingent experience of having to act in response to environmental data. They are called "in-action" because their function is not theorization, formalization, or even explanation. Their purpose is indeed primarily pragmatic: they are attached to classes of situations (in response to which they were constructed) and allow extracting, or rather associating, functional, operational characteristics whose consideration is crucial for the efficiency of the (cognitive and behavioral) activity to be performed on these classes of situations.

Knowledge-in-action first allows segmenting the continuous and formless flow of world information into elementary thought categories; thus selecting, or rather decreeing, the (only) characteristics (taken from the world or rather imposed on it) on which it is deemed pertinent to focus one’s attention in order to consider "what is important from a pragmatic point of view," "what is a source of efficiency." These are what Vergnaud calls “concepts-in-action”, which allow "reading" "retaining" or rather deciding to direct one's gaze on particular dimensions of the world (objects, properties, relations, etc.) that are central to achieving a given objective. As such, each characteristic dimension of the vector space on which a neuron operates, at the input of a neural layer, has a cognitive function of concept-in-action. These dimensions, specific to each layer, allow identifying the characteristics, analytical concepts, categorical abstractions, and thought categories in which it is functional to project, analyze, and categorize the incoming information from this layer, to retain functional aspects relative to the task's purpose.

Knowledge-in-action, Vergnaud tells us, also allows knowing how to coordinate the identified thought categories. And thus, combining them effectively within a coherent modeling of all the retained analytical criteria. This process is carried out by what the researcher calls theorems-in-action, which are local micro-theories, held to be true, relative to the rule by which concepts-in-action interact. As such, a neuron, or more precisely the weight vector defining a neuron, has a cognitive function of theorem-in-action. This is because this weight-vector neuron is a propositional function (f(x,y,z,…)) allowing additive and weighted composition of the dimensions of the input vector space of a given neural layer; that is, selectively fusing some of these dimensions to cognitively act in a particular way on these specific dimensions. Each of these specific cognitive fusions (i.e., each of these neurons) has the effect, in terms of cognitive activity, of generating a new dimension of a new vector space, the output vector space of the layer involved.

The new categorical dimensions of the output vector space of a layer, which are \textit{de facto} those of the input vector space of its successor layer, are in turn, aimed to constitute even more functional concepts-in-action that will themselves then be combined by the neural theorems-in-action of this successor layer into even newer, more abstract, and efficient concepts-in-action.  This is the central cognitive function of a neural network within its consecutive layers: progressively achieving, through a series of steps of increasing conceptual abstractions, a level of segmentation in a categorical vector space whose dimensions have been sufficiently selectively recombined, refined, and made relevant to optimally analyze, within the ultimate neural layer, the information received by the synthetic neural system. The learning phase of a neural network aims to fabricate these successive analytical conceptual stages (i.e., this knowledge-in-action); and its functioning phase then aims to exploit, and apply this iterative conceptual knowledge to ultimately calculate the output (i.e., the output embedding) conceptual values (i.e., dimensional values) that will (ideally) produce the expected cognitive response of this neural network.

\section*{Formal neurons do not decipher  the properties of the world,
they make them emerge by acting cognitively on it
}

Neural concepts-in-action do not epistemologically have an ontological status. These categorical dimensions are not the result of an illusory "decoding" by a formal neural network of pseudo-intrinsic and pre-existing characteristics of a world of properties that would be pre-given and that such a synthetic cognitive system would have the skill to discover, to reveal; and there is indeed cognitive skill, but one of enactive construction in the sense of Varela (1988) and not of revelation in the sense of empirical realism.

The dimensional characteristics mobilized by a neural network are, as we have already mentioned, a function of the architecture and nature of the parameters assigned to it, the singular data with which it was decided to train it, as well as the mathematical operators attributed to it for its learning and functioning. Neural concepts-in-action are thus embodied in the choices of mathematical operations and the structure of these operations that singularly presided over the fabrication of each neural network. As Maturana (1978) and Von Foerster (2003) would say, the vector axes of a given neural layer are not the result of "information transmission"; they pertain not to an "analog copy" but to an authentic "digital reconstruction."

The strength of a neural network is thus not to highlight the \textit{per se} true or good properties of the world's objects; that is, to fabricate representations that would be faithful, adequate mirrors of inherent predicates of the world. But, in a logic of systemic and constructivist circularity dear to Varela, to make functional regularities (in this case, model parameters) emerge during the (cognitive) action of the latter, with its own cognitive-mathematical attributes, on the world's objects. These regularities translate into a bilateral coupling, a mutual shaping, a structural articulation between the world and the synthetic cognitive system that operates its specific thought operations on it.

\section*{Extensions}

We have here limited ourselves to an epistemological reflection concerning the aggregation functions of neural networks. Neglecting for the moment the equally central subjects of (i)  activation functions and their topological effects, (ii)  attention heads and their contextual impacts, and (iii)  different types of neural architectures. These subjects also need to be questioned in an epistemological dynamic.

More broadly, from a perspective of neuro-symbolic hybridization (Alshmrany, 2024; Sun, 2024) of our artificial intelligence technologies, it seems necessary to question the status of a possible direct articulation, at low granularity, of synthetically generated neural knowledge-in-action and formal human symbolic knowledge. Coordination of formal and empirical knowledge, allowing them to mutually regulate and enrich each other from a developmental perspective as mentioned by Vygotsky (1934). Coordination appearing as one of the prerequisites, among others from neuroscience (Minsky, 1988), for the next, radically more advanced generation of artificial intelligence systems.

\section*{Acknowledgements}
The author would like to thank Jourdan Wilson for her meticulous work reviewing and laying out this article.

\end{document}